\documentstyle[twocolumn,aps]{revtex}
\begin{document}
\draft
\title{Defect-unbinding transitions and inherent structures in
two dimensions}
\author{Frank L. Somer, Jr.,$^{1,2,}$\cite{frank} 
G.S. Canright,$^{1,2,}$\cite{email}
and Theodore Kaplan$^3$}
\address{$^1$Department of Physics and Astronomy, University of Tennessee,
Knoxville, TN 37996-1200\\
$^2$Solid State Division, Oak Ridge National Laboratory,
Oak Ridge, Tennessee 37831\\
$^3$Computer Science and Mathematics Division, Oak Ridge National
Laboratory, Oak Ridge, Tennessee 37831}
\maketitle

\begin{abstract}

We present a large-scale (36000-particle)
computational study of the ``inherent structures" (IS) 
associated with equilibrium, two-dimensional, one-component
Lennard-Jones systems.
Our results provide strong support both for the inherent-structures
theory of classical fluids, and for the KTHNY theory of two-stage melting in
two dimensions. This support comes from the observation of {\it three\/}
qualitatively distinct ``phases" of inherent structures: a
crystal,
a ``hexatic glass", and a ``liquid glass". We also directly observe,
in the IS, analogs of the two defect-unbinding transitions (respectively, 
of dislocations, and disclinations) believed to mediate the
two equilibrium phase transitions. Each transition shows up in the
inherent structures---although the free disclinations in the 
``liquid glass'' are embedded in a percolating network of grain
boundaries. The
bond-orientational correlation functions of the inherent structures
show the same progressive loss of order as do the three equilibrium
phases: long-range $\to$ quasi-long-range $\to$ short-range.

\end{abstract}

\pacs{PACS numbers: 61.20Gy,64.70Dv,61.72Bb,61.43Fs}

\section{Introduction}
Some years ago, Stillinger and Weber \cite{SW} introduced a theory of 
liquids, based on the partitioning of the configuration space into 
potential-energy (PE) {\it basins}.  Each of these basins contains a single PE 
minimum, to which all other points within the basin are connected via 
steepest-descent paths.  The PE minima were coined ``inherent 
structures" (IS); all other configurations are taken to be 
vibrational excitations of them.  This approach allows for the 
decomposition of the configurational partition function into a sum,
over PE basins, of intrabasin terms. The resulting partition function may
be approximated as follows:
\begin{equation}
Q=\sum_{\alpha }Q_{\alpha } \approx \int G({\bf p}) Q_{{\bf p}} d{\bf p}
\sim G({\bf p}^{*}) Q_{{\bf p}^{*}}.
\end{equation}
In the first step of this transformation, one splits the partition 
function into a sum over ``basin partition functions" (the usual Boltzmann 
integral, limited to configurations within a given basin), $\alpha$ being 
the basin index. This step is exact in principle.  
The second step transforms this sum into an integral,
via the introduction of the (generally vector-valued) {\it structural 
parameter} ${\bf p}$, characterizing the IS.  
Typically, this parameter would include such
information as average coordination numbers, densities and spatial
distributions of defects, etc. $G({\bf p})$ is a density-of-states
function, enumerating the basins having a given value of ${\bf p}$;
$Q_{\bf p}$ is then the corresponding constrained partition function.
This step in the transformation inevitably loses some information
through the necessity to make $\bf p$ finite-dimensional.
The final step of the transformation is a result of the fact that, in the 
thermodynamic limit, the density-of-states function is essentially 
exponential in the number of particles:
\begin{equation}
G(N,{\bf p})\sim\exp[Ng({\bf p})],
\end{equation}
which permits a maximum-integrand evaluation of the integral over
${\bf p}$.  ${\bf p}^{*}$ is the value of ${\bf p}$ which maximizes
the integrand for a given set of thermodynamic conditions (e.g. volume
and temperature).

The partition function may be further transformed by writing the potential
energy as 
\begin{equation}
\Phi({\bf r})=\Phi_{\alpha}+\Delta_{\alpha}\Phi({\bf r}),
\end{equation}
where $\Phi_{\alpha}$ is the ``structural energy" (the PE of the IS of 
the occupied basin) and $\Delta_{\alpha}\Phi({\bf r})$ is the ``vibrational 
energy" (the difference between the total PE and the structural energy).
This allows the generic basin partition function to be written as
\begin{equation}
Q_{\alpha} = \exp(-\Phi_{\alpha}/k_{B}T) Q^{vib}_{\alpha},
\end{equation}
where
\begin{equation}
Q^{vib}_{\alpha} = \int_{R(\alpha)} \exp(-\Delta_{\alpha}\Phi/k_{B}T) 
d{\bf r}.
\end{equation}
Here, $R(\alpha)$ limits the integration to basin $\alpha$, and $k_{B}$ 
is Boltzmann's constant.  Hence, the total partition function, as given 
by Eq. (1.1), becomes
\begin{equation}
Q = Q^{struct}_{{\bf p}^*} Q^{vib}_{{\bf p}^*},
\end{equation}
where,
\begin{equation}
Q^{struct}_{{\bf p}^*} = G({\bf p}^*) \exp(-\Phi_{{\bf p}^*}/k_{B}T).
\end{equation}
Phase transitions are defined by singularities in the free energy, 
$F_{conf} = -k_{B}T \ln Q$.  Thus, in order for a phase transition to
occur, there must be singularities in $Q^{struct}_{{\bf p}^*}$, 
$Q^{vib}_{{\bf p}^*}$, or both.  The evident unlikelihood of such
singularities, without discontinuities in ${\bf p}^*$ (that is to say
in the types of basins occupied), implies that the existence of marked 
differences between the IS associated with different phases is a 
practical requirement for the applicability of the inherent-structures
theory (IST) to systems 
exhibiting phase transitions.  This requirement has been shown to be
satisfied for certain three-dimensional systems \cite{SW3D}. Also,
some limited results \cite{SW2D} have been obtained for 2D systems.

Recently, we \cite{athens,prl} have performed extensive numerical
studies of inherent structures in simple, single-component fluids in 2D.
An early study \cite{athens} extended the range of such studies to
$N=4096$ particles, but found (as expected) only evidence for two phases
(solid and liquid). More recently \cite{prl} we extended these studies to
$N=36864$ particles. This choice of system size was motivated by earlier
molecular-dynamics studies \cite{Ted,Ted2} giving strong---but not
conclusive---evidence for an intermediate, hexatic phase for systems
of this size and larger (see also Ref.\ \cite{HCA}). Ref.\ \cite{prl}
gave a brief report of the principal results reported here. In this
paper, we offer a detailed discussion of our computational methods and
results. We also provide a clear picture of the disclination-unbinding
``transition" in the IS---a result which was not clear in
\cite{prl}---and some calculations of
the disclination charge-charge correlation function in the equilibrium
fluids. These latter calculations provide further evidence for 
disclination unbinding at the hexatic/liquid transition.

Roughly contemporaneously with the development of IST, Halperin and 
Nelson (HN) \cite{HN}, following work on the melting of 2D solids
by Kosterlitz and Thouless \cite{KT}, 
predicted a two-stage melting mechanism for 2D systems.  
A number of results on the first stage of melting were obtained
independently by Young \cite{Y} (see also Nelson \cite{N}).
In the resulting picture of two-stage melting---commonly called the `KTHNY 
theory'---each successive phase (in order of increasing energy) is 
characterized by the presence of an additional type of defect:  the 
solid contains only dislocations, bound in pairs; the intermediate, 
{\it hexatic\/} phase adds unbound dislocations; and the liquid further 
adds unbound disclinations.  Attendant to this progression of defects are 
differences in the bond-orientational correlation function, which exhibits 
long-range order, power-law decay, and exponential decay 
for the solid, hexatic, 
and liquid phases, respectively.  (For a detailed review of the KTHNY 
theory and the defects involved, see the review of Strandburg 
\cite{Strandburg}.)  Since KTHNY predicts the existence of {\it three\/} 
condensed phases---for which IST requires an equal number of distinct 
classes of IS---2D would seem to offer an ideal testing ground for IST.  
Furthermore, it seems that a study of the IS underlying the different phases 
in 2D systems might provide novel and useful microscopic evidence for
the defect-unbinding transitions expected from the theory---assuming
that such defects can be `trapped' by the quenching procedure (which
yields the mechanically stable inherent structure from a snapshot
configuration at thermal equilibrium).
Indeed, the defects present in 
each equilibrium phase should also show up in the IS underlying that 
phase---but much more clearly, due to the attendant removal of the 
vibrational distortions present at equilibrium---{\it if\/} these defects 
are mechanically stable. Before discussing our numerical results,
then, we will discuss the question of the mechanical stability of 
the defects---dislocations and disclinations.

The interaction energies of the defects are typically calculated by well 
known methods of linear elasticity theory \cite{Nabarro}.  To make an 
elastic defect, one takes an unstrained elastic medium and introduces
one or more topological changes---yielding one or more topological
defects---by cutting, shifting, and gluing.
The next step is to invoke the ``equations of equilibrium" (EOE)---which 
require that there be no net force at any point in the medium---{\it at 
constant topology}.  This then gives the strain field 
which, in turn, gives the self-energies and energies of interaction for 
the defects.  The assumption of constant topology is crucial, because 
it is the topology that defines the elastic problem to be solved.
However, it is important to note that this can {\it only\/} be an
assumption in the continuum theory: in the absence of a microscopic atomic
structure in the fluid, there is no reason to expect any dislocation
to be pinned (against the calculated forces of attraction or repulsion) 
at any point in the medium, except `by hand'. 

Thus we cannot expect to see such structures as free dislocations in 
any {\it mechanically stable\/}
configuration,
unless some
justification can be given for this assumption of constant topology.  
For the case of dislocations, this justification comes in the form of 
the Peierls-Nabarro potential \cite{Nabarro}.  This is a periodic 
``corrugation" in the inter-dislocation potential, arising from the 
underlying microscopic structure of the material. This potential 
is well known to 
be capable of pinning dislocations, such that arrays of dislocations may 
be rendered mechanically stable. This leads us to anticipate
that the configurations of dislocations trapped in our
numerically-obtained IS may in fact provide useful insight into the
equilibrium defect structures, {\it without\/} the almost overwhelming
`noise' associated with the vibrations about the IS, occurring
in thermal equilibrium.

There is less justification for this assumption, as applied to disclinations.
In fact, there is reason to doubt the mechanical stability---and hence the
presence in IS---of free disclinations \cite{Nabarro}.  
The question is then, can the inherent-structures idea, invoking as it
does a qualitative difference in IS between different thermodynamic
phases, be reconciled with the KTHNY picture of melting (hexatic $\to$ liquid)
by the unbinding of disclinations---even when there is good reason to
expect that no free disclinations can be seen in mechanical equilibrium
(i.e., in any IS)? We provide a conclusive (`yes') answer to this question,
below, while at the same time failing to find any evidence for
mechanical stability of free disclinations.

Our results reveal an extremely clean correspondence between the
predictions of the KTHNY theory of two-stage melting and the 
inherent structures associated with each thermodynamic phase.
Previously, the principal barrier to this sort of study has been
the difficulty associated with finding the hexatic phase in simulations 
\cite{simulations}, in part due to limitations in system size.  
Boundary conditions and equilibration methods may also play an important
role.
The simulations in Refs.\ \cite{Ted} ($N>100000$) and \cite{HCA} 
($N\sim 65000$) gave 
some compelling evidence for the hexatic phase. However this phase was
found to be only {\it metastable\/} thermodynamically in \cite{Ted};
and the differences in method and boundary conditions in the two studies
leaves some room for controversy. Our own studies use a system size
($N\sim 36000$) for which a metastable hexatic phase appeared in
the study of Ref.\ \cite{Ted}, with quenches from the `equilibrium' 
hexatic phase being
taken from snapshots in this metastable thermodynamic state.
We believe that our results, revealing as they do three classes 
(one clearly `hexatic') of
IS for these fluids---classes which are expected to persist in the
thermodynamic limit---provide further support for the hypothesis that
the hexatic phase is thermodynamically stable, in some region of the
phase diagram, for $N\to\infty$. This evidence is distinct from, and
complementary to, that obtained from equilibrium studies. Our results
also further strengthen the basic premise of inherent-structures theory:
that distinct thermodynamic phases are characterized by qualitatively
distinct inherent structures, such that singularities in thermodynamic
functions may be ascribed to discontinuities in the occupation 
probabilities of potential-energy basins, at a phase boundary.

\section{Computational procedure}
In order to study inherent structures, one first needs starting 
configurations, taken as snapshots from thermal equilibrium. 
In the present study, these were obtained 
directly from the molecular-dynamics (MD) simulations
described in Ref.\ \cite{Ted}, which made use of 
a computational framework described by Melchionna, Ciccotti, and Holian 
(MCH)\cite{NPTMD}---specifically, a constant-{\it NPT} molecular dynamics 
simulation, using the Parrinello-Rahman \cite{PR} shape-varying box
with periodic boundary conditions (BC).
Let us, then, give a brief summary of those aspects of this framework 
which are most relevant to the present study, the details being available 
in Ref.\ \cite{NPTMD}.  

For the sake of simplicity, we will outline the method for the case of 
isotropic volume fluctuations and then state the changes necessary to 
account for the shape-varying box.  The MCH equations of motion, for
the case of isotropic volume fluctuations, are
\begin{eqnarray}
\nonumber & \dot{\bf r}_{i}=\frac{{\bf p}_{i}}{m_{i}}+\eta({\bf r}_{i}-
{\bf R}_{0}) \\
\nonumber & \dot{\bf p}_{i}={\bf F}_{i}-(\eta + \zeta){\bf p}_{i} \\
& \dot{\zeta}=\nu_{T}^{2}\left[\frac{T(t)}{T_{ext}} - 1\right] 
\label{MD_eq} \\
\nonumber & \dot{\eta}=\frac{\nu_{P}^{2}}{Nk_{B}T_{ext}}V[P(t) - P_{ext}] \\
\nonumber & \dot{V}=dV\eta.
\end{eqnarray}
Here, ${\bf r}_{i}$, ${\bf p}_{i}$, and ${m_{i}}$ are the position, momentum,
and mass, respectively, of particle $i$.  ${\bf F}_{i}$ is the instantaneous 
force acting on particle $i$, and ${\bf R}_{0}$ is the center of mass of the
system.  $\eta$ is a barostating factor which tends to restore the 
instantaneous pressure, $P(t)$, to the preset value $P_{ext}$.  It is 
modulated by the adjustable parameter $\nu_{P}$, which is termed the 
``barostating rate".  Similarly, $\zeta$ serves to equilibrate the 
temperature and is tuned by way of the ``thermostating rate", $\nu_{T}$.  
$V$ is the volume and $d$ is the dimensionality.  The main alterations to 
Eq.~(\ref{MD_eq}), needed to accommodate a shape-varying box, are to change 
the scalars $\eta$ and $P$ to tensors and to replace $V$ with a ``box 
matrix" whose columns are the basis vectors of the box.  The basic form of 
the equations remains that of Eq.~(\ref{MD_eq}).  The potential used was of 
the shifted Lennard-Jones form,
\begin{equation}
V(r)=\left\{ \begin{array}{lcccccccc}
4 \epsilon [(\sigma/r)^{12} - (\sigma/r)^{6}] + V_{c}, &&&&&&&& r<r_{c}, \\
0, &&&&&&&& r \ge r_{c}, \end{array} \right.
\end{equation}
where $\epsilon$ and $\sigma$ are parameters, $r$ is the interatomic
distance, $r_{c}$ is the cutoff radius, and 
\begin{equation}
V_{c}=-4 \epsilon [(\sigma/r_{c})^{12} - (\sigma/r_{c})^{6}].
\end{equation}
In units in which $\epsilon$, $\sigma$, $k_B$, and $m$ (the atomic mass)
are all equal to 1---units which we use throughout this paper---the 
parameters used in the equilibrations were as follows:
\begin{equation}
  \begin{array}{l}
    N=36864 \\
    P_{ext}=20 \\
    T_{ext}\ \left\{ 
      \begin{array}{ccl}
        \leq 2.15&&{\sf (crystal)} \\
        = 2.154&&{\sf (hexatic)} \\
        \geq 2.17&&{\sf (liquid)}
      \end{array}
    \right. \\
    r_{c}=4 \\
    \Delta t = .0005,
  \end{array}
\end{equation}
where $N$ is, of course, the number of particles.

Having obtained an equilibrium configuration (in the hexatic case, this is
only a thermodynamically {\it metastable\/} equilibrium \cite{Ted}), 
in order to find the associated IS one must perform a steepest-descent
minimization of the total potential energy---which, for the current 
constant-$P$ case, consists of the total interparticle potential plus 
$P_{ext}V$.  In practice, this is too time consuming to be practical 
\cite{SW3D} for the large system size of the current study---indeed, it is 
impractical for all but the smallest systems.  Instead, we make use of the 
above-outlined MD method.  Beginning with an equilibrium configuration, we 
first zero all particle velocities and the velocities of the ``box walls".  
We then run the MD simulation at very low temperature ($T_{ext}$ 
something like $10^{-6}$ to $10^{-4}$ of the equilibrium temperature), 
carefully adjusting the thermostating and barostating rates, such that the 
instantaneous temperature remains very close to $T_{ext}$ and the PE 
smoothly decreases with the time.  Furthermore, the PE is checked at each 
time step, and if an increase is found, we go back to a previous 
configuration (saved before the occurrence of the increase), re-zero the 
velocities, and restart the simulation.  This process is continued until 
the duration of the MD runs (i.e.\ before a PE increase occurs) becomes only 
a few time steps.  At this point, we run the MD simulation at the same, 
very low temperature (and without, of course, the requirement of a strictly 
decreasing PE), long enough (typically something like $10^5$ time steps) to 
ensure that the system is, indeed, vibrating about a PE minimum.  If this 
test is successful, we have an IS.  (At such low temperatures, the 
vibrations are of small enough amplitude to be negligible for structural 
considerations.)  Otherwise, we continue the minimization procedure, until 
a minimum is found which does pass our test.

Of course, the structures obtained from the above-outlined minimization 
procedure (hereafter referred to as ``quenching") will, in general, differ 
from the `true' IS, connected by a steepest-descent path to the starting 
equilibrium configuration.  However, the very low temperatures---which is 
to say very small particle momenta---maintained throughout the quenching
process ensure that the system trajectory nearly \cite{nearly} follows the 
steepest-descent path prescribed by IST.  Also, according to IST, fluid 
systems at equilibrium transit frequently and {\it exclusively\/} 
(at least for $N\to\infty$) between 
{\it thermodynamically equivalent\/} PE basins \cite{Somer} (i.e.\ basins 
having the same value of the structural parameter, ${\bf p}$).  
Hence we believe that the structures obtained by our quenching procedure 
will be thermodynamically equivalent to the actual IS of the starting 
configuration.  That is, they should be representative of the set of 
IS associated with the thermodynamic conditions of the equilibrium 
configuration. 

\section{Results}
In earlier work \cite{athens} we performed a large number of quenches of
equilibrium systems at $N=4096$. In the present work we increase $N$ to
36864 particles, in order to be able to quench from all three thermal
phases: solid, hexatic, and liquid. At this system size the quenches are
very intensive computationally. We have performed three quenches---following
the procedure outlined above---for each phase, and several additional
quenches which did not strictly enforce the requirement of monotonic
decrease of the PE. The results were qualitatively the same for all
quenches derived from the same starting phase.

Fig.\ 1(a) shows an equilibrium snapshot for the solid phase.
We plot only those atoms which are not 6-fold coordinated according to a
Voronoi construction, labeling all such atoms by their coordination
number (mostly 5 or 7). Fig.\ 1(b) then shows the configuration of
Fig.\ 1(a), when relaxed to {\it mechanical\/} equilibrium by our
quenching procedure. Although Fig.\ 1(b) is mostly white space (i.e.,
6-fold coordinated atoms), we include it here to illustrate
the dramatic reduction in defect number as a result of quenching from
thermal equilibrium to mechanical equilibrium. It is apparent to the eye
that there are no free dislocations in the solid-derived IS: every
dislocation is closely bound in a (vector-)charge-neutral composite.
We have also computed the bond-orientational correlation functions (BOCF)
for the various quenches. Obviously, for the solid-derived IS, the BOCF
has long-ranged order.

On quenching from the (metastable) equilibrium hexatic 
phase, we obtain structures such as that shown in Figure 2.  
(We do not show the equilibrium defect configuration as the defects are
very dense.)  Clearly, there is a large density of defects, 
{\it even in mechanical equilibrium}, for this case.
Of course, there are still some bound dislocations, some of which compose 
large-angle grain boundaries identifiable as chains of very closely spaced 
dislocations. In addition to these---and in contrast to the crystal 
IS---there are many dislocations which do not have any `canceling' 
dislocations within several lattice spacings, some of which show a clear 
tendency \cite{FHM} to arrange themselves into small-angle grain boundaries.
We term these the ``free'' dislocations for the IS; reasoning from the 
existence of the Peierls-Nabarro potential, we argued above that free 
dislocations, {\it if\/} present in the equilibrium snapshot, will survive 
the quench and thus appear in the IS. Here, we claim that a comparison of 
Figs.\ 1 and 2 graphically reveals the dislocation-unbinding transition 
{\it in the inherent structures}.

We can also test this idea with the BOCF.
While networks of large-angle grain boundaries are capable 
of destroying the quasi-long-range orientational order characteristic of 
the hexatic phase, those present in our hexatic quenches are relatively 
small and isolated, so that this order is in fact preserved.  
Log-log plots of the BOCF [$g_{6}(r)$] for a 
typical hexatic ``MD snapshot" \cite{Ted} 
and its associated quenched structure \cite{prl} reveal that both obey a
power-law behavior, with the IS showing a smaller exponent (i.e.\ a 
slower rate of algebraic decay of the orientational order). This may be
attributed to the removal, on quenching, of long-wavelength torsional 
phonons, which are supported by the hexatic phase's finite Frank constant.

An IS of this nature has not been seen in any smaller system. 
In fact---if we use
the word ``glass'' as shorthand for structural glass, i.e.\ an atomic
configuration in mechanical (but not thermal) equilibrium---then 
Fig.\ 2 shows a hexatic glass. Two-dimensional glasses have mostly been
studied using two or more atomic species
\cite{2Dhexglass,2D2comp,bearings}, since the 
`frustration' in 2D monatomic fluids is very small (it is zero for the
2D packing problem) \cite{frustration}. Hexatic glasses have, to our
knowledge, only been seen before in two-component systems---in
simulations \cite{2Dhexglass}, and in ball-bearing experiments
\cite{bearings}.

With the isotropic liquid as the starting point, our quenching procedure 
results in structures such as that shown in Figure 3.  The two-stage
melting theory predicts that the transition from hexatic to isotropic
liquid takes place via the unbinding of disclinations (our 5's, 7's
etc.). However there are no free
disclinations in any of our liquid-derived IS. 
Rather, the only additional defects, as compared 
to the hexatic phase, are---as is clear from Fig.\ 3---percolating 
networks of large-angle grain boundaries.  These in themselves can
destroy the quasi-long-ranged orientational ordering, as may be verified
by calculating the BOCF for the IS of Fig.\ 3. In fact, both the
equilibrium snapshot and its IS show an exponential decay of
orientational order, with roughly the same exponent \cite{prl}.

It is thus tempting to suppose that the equilibrium liquid is also
characterized by percolation of grain boundaries---i.e., that the hexatic
$\to$ liquid transition takes place by grain-boundary
melting \cite{Chui}. Certainly [compare Figs.\ 2(a) and 3] the 
transition appears in the IS as a percolation of grain boundaries.
However, we believe that we {\it can\/} see the unbinding of 
disclinations in our IS, with a bit more effort.

In Fig.\ 4 we show the quenching of an artificial starting condition
whose only defects are four widely
spaced disclinations (two positive and two negative).
The corresponding quenched (mechanically stable)
structure is a roughly square grain-boundary network, whose nodes 
correspond closely to the positions of the original disclinations.  
The ``free'' disclinations of Fig.\ 4(a) are (as expected)
not mechanically stable; and they relax upon quenching to a
network of grain boundaries, which serves as a ``fossil relic''
of the free disclinations in the starting configuration.
This suggests that there should be
a strong correlation between the average separation of free
disclinations in an equilibrium configuration, and the average grain size 
in the corresponding quenched structure.  We find further support for
this idea from other quenches like that shown in Fig.\ 4: above a
threshold separation distance, the grain size in the quenched structure
closely reflects the spacing of the original disclinations. (For
disclinations closer than the threshold distance, the relaxed structure 
is a single grain.) Thus, the fact that the 
equilibrium and quenched configurations have nearly identical orientational
correlation lengths \cite{prl} is consistent with the presence of free 
disclinations in the equilibrium liquid.

It is also interesting to note that, while the transformation shown in 
Fig.\ 4 is quite dramatic, the final configuration [Fig.\ 4(b)] still, in
a sense, contains ``free" disclinations.  These are at the nodes of the 
grain-boundary network, which are near the positions of the 
original disclinations of Fig.\ 4(a).  Disclinations are defined as centers
of lattice rotation; that is, tracking the local lattice 
orientation, while making a closed circuit around a disclination, will show 
a net rotation.  For the present case, this means that if we track the 
orientation of six-coordinated cells, as we make a closed circuit around
a disclination, we will find a net rotation of some integral multiple of 
$\pi/3$.  The disclinations centered on the five- and seven-coordinated 
atoms of Fig.\ 4(a) give rotations of
$+\pi/3$ and $-\pi/3$,
respectively.
Similarly, on making a circuit enclosing a set ($A$) of atoms whose 
average coordination number is different from six, we will find a net 
lattice rotation---specifically, 
\begin{eqnarray}
\nonumber & \theta_{rot.}=\frac{\pi}{3}q_{A} \\
& q_{A}=\sum_{i \in A}(6-z_{i}).
\end{eqnarray}
Here, $q_{A}$ is the net ``disclination charge" 
in $A$, and $z_{i}$ is the coordination number of atom $i$.
Of course, in order to properly define this lattice rotation, we need a
circuit consisting solely of ``good crystal" (i.e.\ six-coordinated cells).
Such circuits do exist around the grain-boundary nodes of Fig.\ 4(b) 
(as shown
in closeup in Figure 5). Hence we see that there are, indeed, ``net-sevens"
($q_{A}=-1$) and ``net-fives" ($q_{A}=1$) at the grain boundary nodes, 
near the positions of the original negative and positive disclinations, 
respectively.  

We next examine the distribution of such ``net disclinicity" in our IS.
In doing this, the disclinations are identified as
outlined above---by identifying groups of atoms whose average coordination
number is different from six, and which can be enclosed by a path consisting
entirely of six-coordinated atoms.  If, for a given disclination, the 
smallest such path encloses more than one atom, the location of the 
disclination is somewhat arbitrary (except, of course, that it should be 
somewhere within the enclosed area).  For the purpose of illustration in 
the present work, we have used the following rule:  if the disclination is 
near a grain-boundary node, we assign it to the appropriately coordinated
atom ($z=5$ for a positive disclination, or $z=7$ for a negative 
disclination) nearest the node; otherwise, we assign it to the 
appropriately coordinated atom nearest to the closest, opposite-signed
disclination.  Application of this procedure to the quenched structures of 
Figs.\ 2 and 3 results in the disclination arrays shown in Figure 6.
Comparison of these arrays with the structures from which they were 
obtained reveals that the presence of {\it net\/}
disclinations is restricted to the vicinity of large-angle grain boundaries.  
That is, Figs.\ 2, 3, and 6, considered together, are entirely
consistent with the notion obtained from Fig.\ 4: that free
disclinations, upon quenching, relax to mechanically stable IS in which a
network of grain boundaries marks the extent, and even positions, of the
original distribution of free disclinations. In this view, then, the
grain-boundary-percolation transition seen in our IS is the direct 
analog of the disclination-unbinding transition in equilibrium.
In other words, if the disclinations unbind, the grain boundaries in the
IS {\it must\/} percolate; and, on the other hand,
bound disclinicity appears in the IS as localized (or no) grain boundaries.
Our net-disclination algorithm then simply erases the grain-boundary
network, revealing bound disclinations in our hexatic IS [Fig.\ 6(a)]
and unbound disclinations in the liquid IS [Fig.\ 6(b)].

Thus we believe that we have seen {\it both\/} defect-unbinding
transitions reflected in our IS. However this conclusion requires 
a chain of reasoning which is not airtight.
Hence, as a further test of the hypothesis that disclination unbinding
mediates the hexatic $\to$ liquid transition, we have
analyzed the disclination distribution
of the two {\it equilibrium\/} phases, using the disclination ``charge-charge" 
correlation function (CCF):
\begin{eqnarray}
\nonumber & g_{q}(r)=\sum_{i \neq j}\delta(r-r_{ij})q_{i}q_{j} \\
& r_{ij}=|{\bf r}_{i} - {\bf r}_{j}| \\
\nonumber & q_{i}=6-z_{i},
\end{eqnarray}
where the sum is over all pairs of atoms.  According to Halperin 
\cite{Halperin}, the absolute value of this function should exhibit 
(asymptotically) a power-law decay when the disclinations are bound
(but interacting with a logarithmic potential, as in the hexatic phase),
and an exponential decay for the case of unbound disclinations.
As suggested by 
Fig.\ 6, typical equilibrium configurations contain relatively few 
candidates for free disclinations.  Additionally, at equilibrium there 
are very many non-sixfold-coordinated atoms that do not comprise unbound 
disclinations---they are components of dislocations, both stable and 
``virtual"---so that the disclination CCF must be extracted from the 
considerable noise resulting from these other defects.  For these reasons, 
the elucidation of the asymptotic behavior of the CCF requires
long-time averaging over very many equilibrium configurations.  
This is an extremely time consuming process. We have obtained
results which, we believe, offer some evidence for the unbinding of
disclinations in the liquid phase. However, even after much time averaging,
our CCFs show meaningful behavior only over a single decade of
distance; hence we cannot claim that this evidence itself is
conclusive.

Our CCF results for 36K particles are presented in Figure 7 (hexatic
phase) and Figure 8 (liquid). The severe noise in these data is clearly
evident. However we also believe that there is clear visual support in
these data for the hypothesis that the hexatic CCF is better fit by 
a power-law or subexponential form, while the liquid CCF is better fit
by an exponential. Obviously one can only draw very tentative
conclusions from these data. However, from our experience we believe that 
further time averaging will not significantly reduce the noise in these
data; instead, larger systems (with larger numbers of defects) need to
be examined.

\section{Discussion}

Our principal conclusion from these studies is that the basic premise of
inherent-structures theory is well confirmed for 2D simple fluids.
That is, inherent structures obtained from differing equilibrium
phases differ from one another in qualitative and reproducible ways.
Our own results extend those from previous IS studies of 2D fluids
\cite{SW2D} in two ways. We have studied system sizes 1--2
orders of magnitude beyond those of previous studies. This increase in
size has enabled us to obtain, for the first time, convincing
differences among {\it three\/} distinct phases in 2D.

In particular, we find unambiguous evidence for the existence of {\it
hexatic\/} inherent structures, which are also hexatic (structural)
glasses. Such structures do not appear at the level $N=4000$
\cite{athens}, but are found at $N=36000$---at which size the equilibrium
hexatic phase is thermodynamically metastable \cite{Ted}.
It is certain that the hexatic IS will persist for all larger $N$.
The question is then (in the language of IST), can these IS reach
a balance of entropy and energy (with increasing $N$)
such that the hexatic phase becomes a true 
minimum of the free energy, in some part of the phase diagram?
Beyond the minimal criterion of showing the existence of such IS, our
present results do not answer this question. However, more detailed 
studies of the IS reported here may allow further progress.

Our results are also relevant to the KTHNY theory of two-stage melting.
Besides the fact that we have shown three ``phases'' of inherent
structures---with the same types of bond-orientational order as the
corresponding equilibrium phases---we have also obtained graphic
evidence for both defect-unbinding transitions. The unbinding of
dislocations, as one moves from solid-derived to hexatic-derived IS, 
is apparent. In contrast, free disclinations are (apparently) not
mechanically stable [Fig.\ 4], and so hidden in the liquid-derived IS. 
However, we have used a simple,
deterministic algorithm which identifies and reveals net disclinicity
(i.e., that not canceled by neighboring atoms) in any IS. 
This rule, applied to our hexatic- and liquid-derived IS,
yields configurations of defects showing
a clear disclination-unbinding transition. This latter transition
shows up as a percolation transition for grain boundaries in the
untransformed IS; the net disclinations are then (typically) found at
the nodes of the grain-boundary network. We have also obtained some
further evidence for disclination unbinding by studying the disclination
charge-charge correlation function for the equilibrium fluid.

Finally, we reiterate that inherent structures are, in principle,
structural glasses: nonequilibrium (and disordered) configurations 
trapped away from the equilibrium state by potential-energy barriers.
Hence we find that there are {\it two distinct types\/} of structural
glasses for one-component, Lennard-Jones fluids in 2D. These two types 
might be experimentally observable---although it seems likely that 
the required quench rates are too high. The idea however has
considerable interest. Just as the novel phases found in 2D equilibrium
phase diagrams have enhanced our knowledge of the phases of fluid matter
in general, so may unusual structure in a 2D ``glassy phase diagram''
be expected to offer new insights into our general understanding of
glassy matter.

{\it Acknowledgments.}--
We thank David Nelson for helpful comments regarding free disclinations,
and Ken Stephenson of the 
University of Tennessee's Mathematics Department for generating starting 
structures such as that shown in Fig.\ 4(a).  This work was supported in 
part by the Applied Mathematical Science Research Program, Office of
Energy Research, and the Division of Materials Science,
U.S. Department of Energy, through Contract No. 
DE-AC05-84OR21400 with Martin Marietta Energy Systems Inc.  FLS and GSC 
were supported by the NSF under Grant \# DMR-9413057.

\begin{figure}
\caption{
(a) A snapshot configuration taken from the solid phase in thermal
equilibrium; the particle number is $N=36864$ and the (dimensionless)
snapshot temperature is $T_s = 1.988$. Only the positions of atoms which
are not sixfold coordinated are shown. The outer parallelogram shows the
periodic boundaries of the deformable 2D box in which the MD simulations 
take place.
(b) The inherent structure obtained from (a) by `quenching'---i.e.,
(nearly) steepest-descent minimization of the potential energy $U+P_{ext}V$
[where $U$ is the interaction energy, $P_{ext}=20$ the external
pressure and $V$ the 
(variable, two-dimensional) volume].  The reduction in defect number from 
(a) is dramatic. Although it is not possible here, close examination of the
defects in (b) shows that there are no dislocations that are not closely
bound into neutral composites.
}
\end{figure}

\begin{figure}
\caption{
(a) Inherent structure for 36864 particles with periodic BC,
obtained by relaxing a configuration from a hexatic phase in (metastable)
equilibrium. The relaxation is done at constant pressure $P_{ext} = 20$,
from an equilibrium snapshot at $T_s=2.154$.
Again, only those atoms which are not 6-fold coordinated are marked.
(b) Enlargement of the boxed-in section of (a).  Free dislocations appear
as 5-7 pairs which are neither bound into neutral composites, nor
in
large-angle
grain boundaries.}
\end{figure}

\begin{figure}
\caption{Inherent structure obtained by relaxing an equilibrium liquid
configuration. The parameters are as in Figs.\ 1 and 2, except $T_s=2.17$.
The 
large-angle
grain boundaries, isolated in Fig.\ 2, span the sample here, and 
in all other liquid quenches we have done. Examination at finer scale
shows no
free disclinations (which would appear here as isolated 5's and 7's).}
\end{figure}

\begin{figure}
\caption{
(a) An artificial starting configuration for 4096 particles, constructed
with 4 widely spaced disclinations of zero net scalar and vector charge.
Again we use periodic BC.
(b) The relaxed structure for (a) (defects only). The free disclinations
have vanished; what remains is a network of grain boundaries which 
closely marks the original positions of the disclinations.
}
\end{figure}

\begin{figure}
\caption{Enlargements of the (a) upper-left and (b) lower-left 
grain-boundary nodes of the quenched structure of Fig.\ 4(b), 
corresponding to positive and negative disclinations, respectively.  
Six-coordinated atoms are marked by asterisks, while non-six-coordinated 
atoms are marked by their coordination numbers.  The arrows are an aid to 
the eye, showing the lattice rotation on making a circuit around the 
clusters having a net disclination charge.}
\end{figure}

\begin{figure}
\caption{The structures of (a) Fig.\ 2(a) and (b) Fig.\ 3, showing only
the `net' disclinations, as indicated in Fig.\ 5, and 
located by the method described in the text.  Each
positive disclination is marked by a ``5", and each negative disclination
is marked by a ``7". (a) We see that net disclinicity in the inherent
structure is confined to the region of large-angle grain boundaries.
For the hexatic IS this region does not span the sample. (b) For the
liquid IS, both the grain boundaries [Fig.\ 3] and the net disclinations
(shown here) span the system. Hence (a) and (b) here show the
inherent-structures analog of disclination unbinding.
}
\end{figure}

\begin{figure}
\caption{The disclination charge-charge correlation function 
[$g_q(r)$,  or ``CCF'', as defined
in the text], obtained for the equilibrium hexatic phase (36K particles)
at $T=2.154$;
(a) semilog; (b) log-log plot. 
The distance $r$ is in units of the Lennard-Jones parameter $\sigma$,
in this figure and in Fig.\ 8.
There is a range, approximately
$2 < r < 20$, which is not completely dominated by finite-size artifacts. 
In this range, the hexatic CCF is roughly linear on the log-log plot (b), and
hence concave upwards in (a).
}
\end{figure}

\begin{figure}
\caption{The disclination CCF obtained for the liquid at $T=2.327$;
(a) semilog; (b) log-log. Again there is approximately one decade (in
$r$) of data not dominated by noise. Here the CCF is roughly linear on
the semilog plot (a), and so convex upwards on the log-log plot (b).
}
\end{figure}


\begin{references}
\bibitem{nearly} It is plausible, given the expected `rugged' topography
of the PE surface, that both a steepest-descent minimization, and our 
own quenching procedure, may be in fact chaotic---i.e., showing sensitive
dependence on initial conditions---for large enough $N$. However `close'
descent paths (that is, those distinguished only by small differences in
initial conditions) may still be expected to yield thermodynamically
equivalent inherent structures.
\end{references}
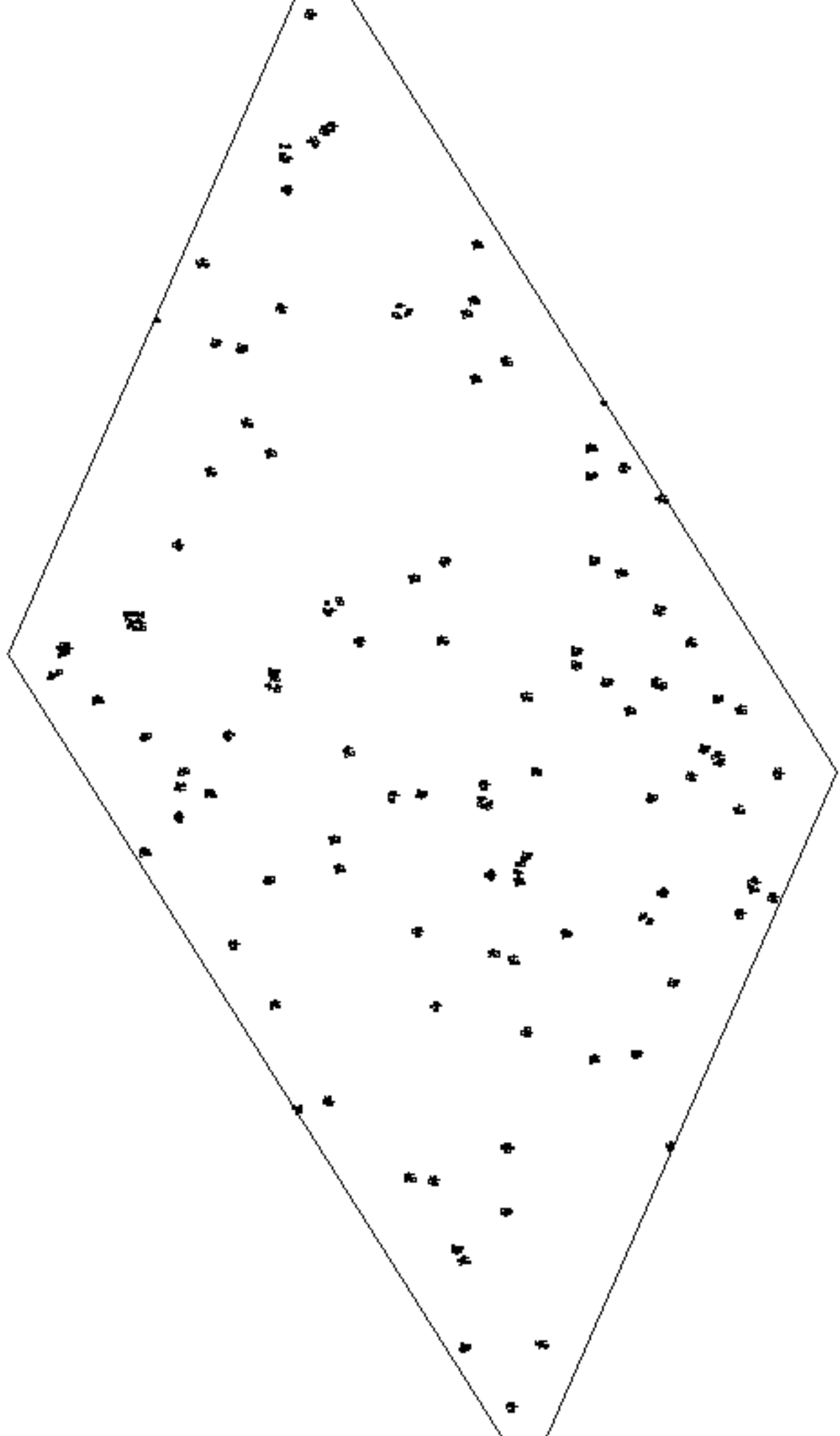
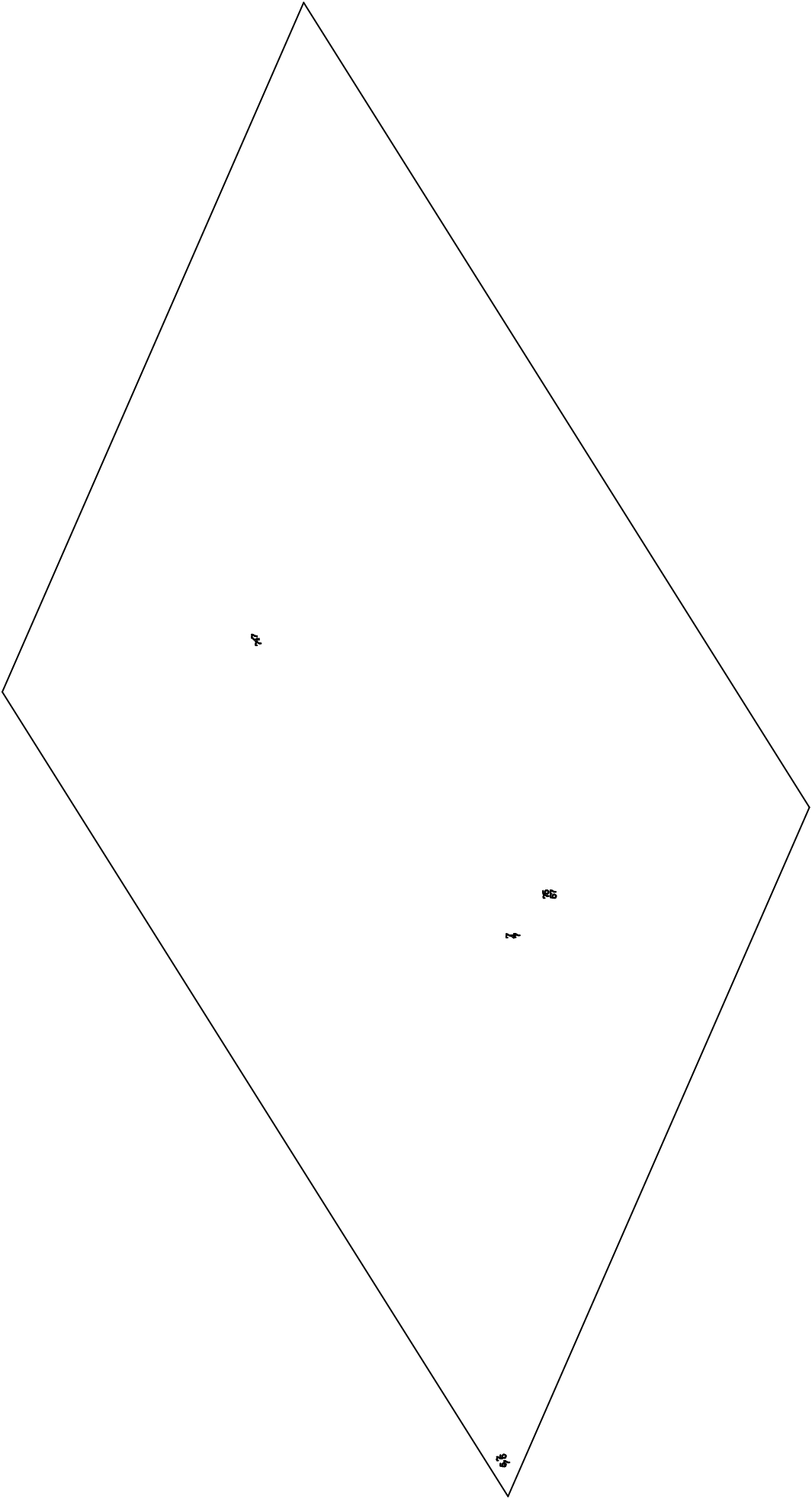
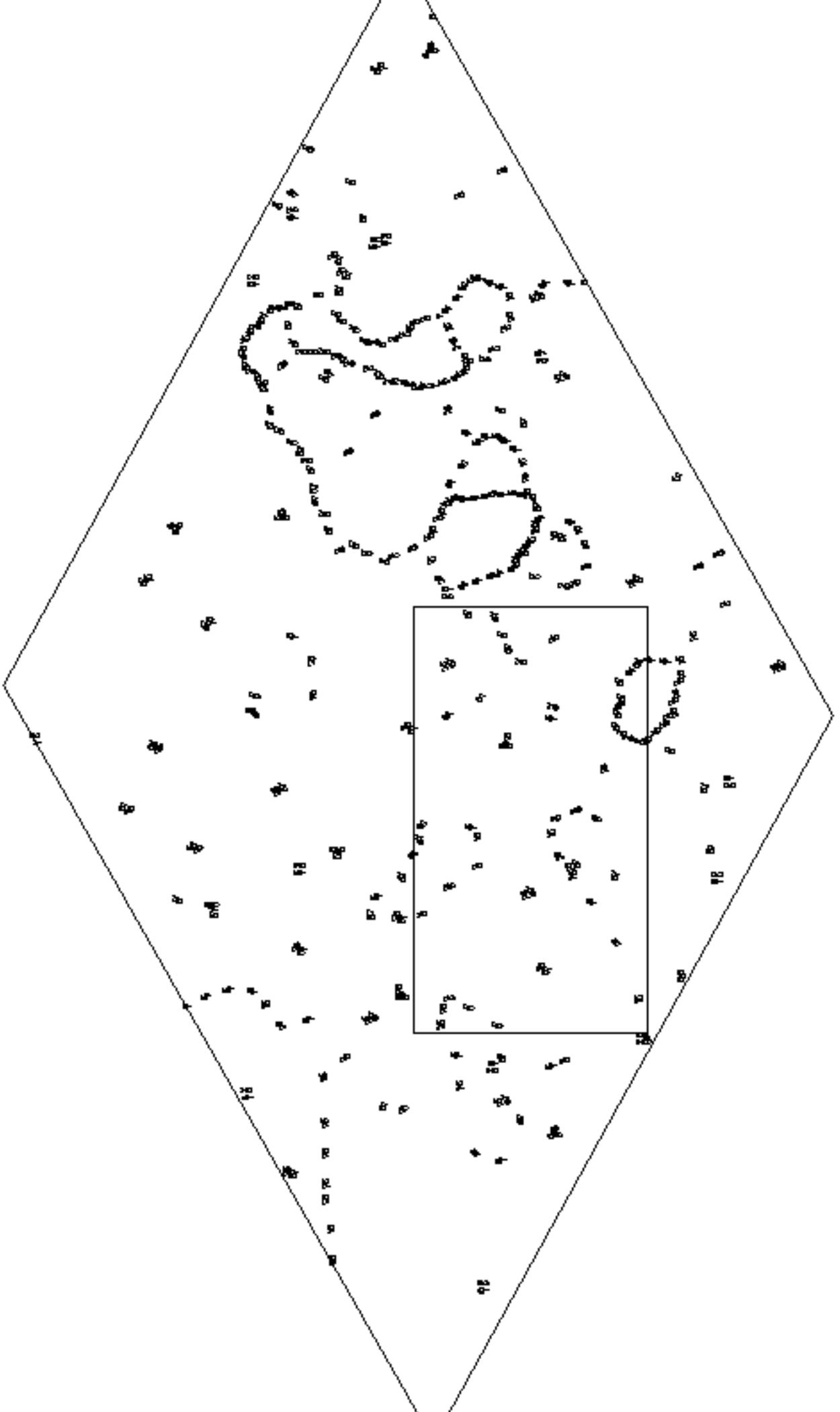
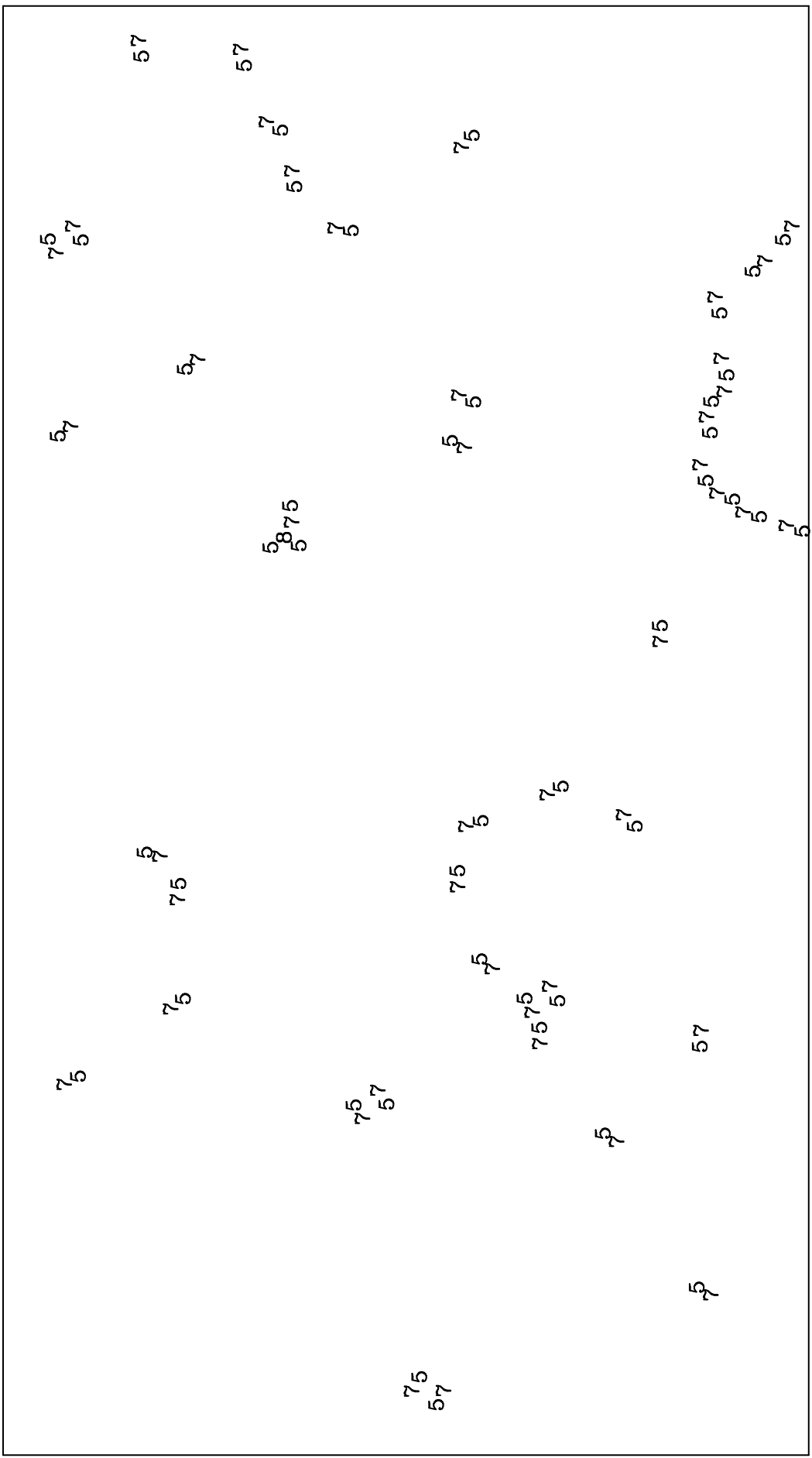
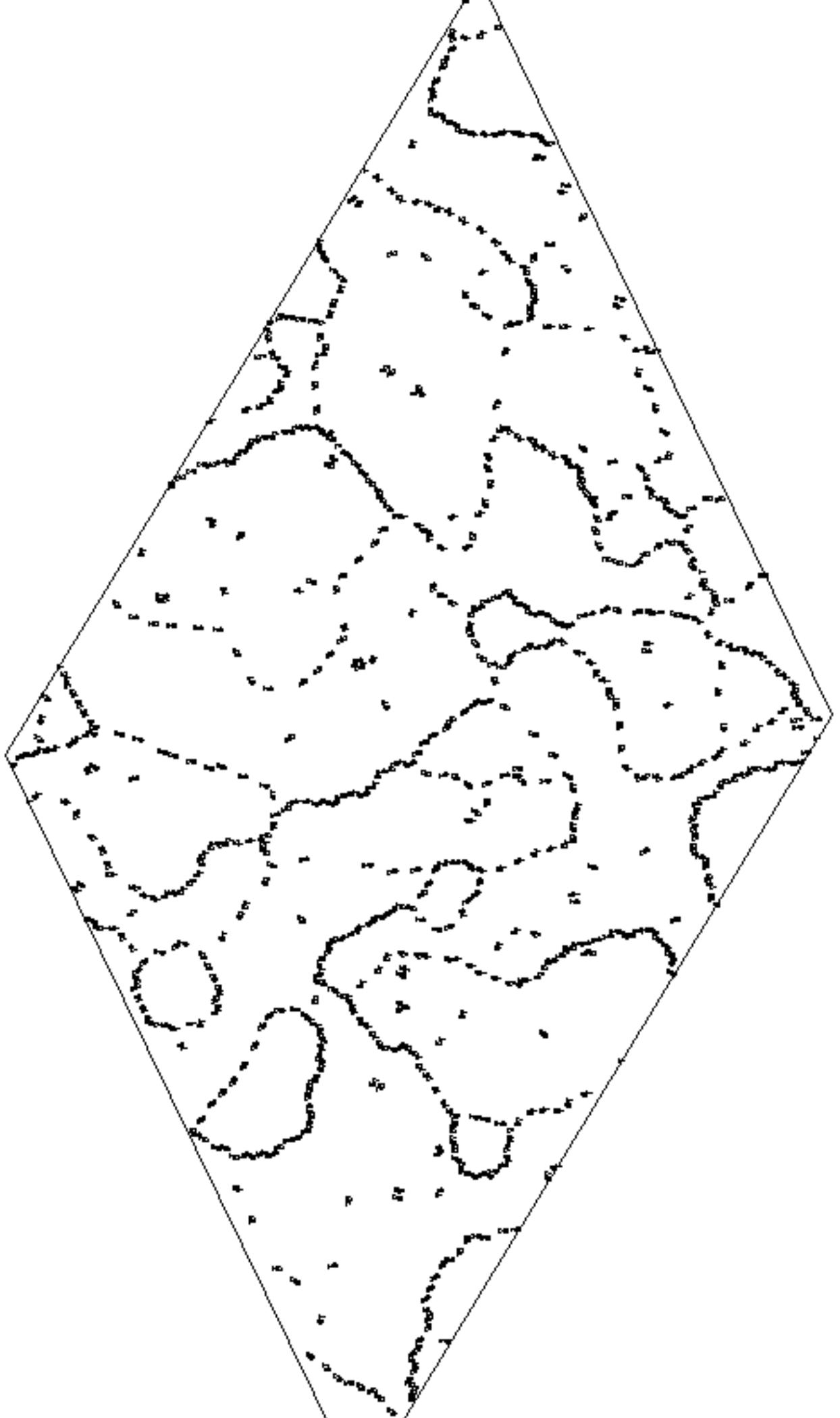
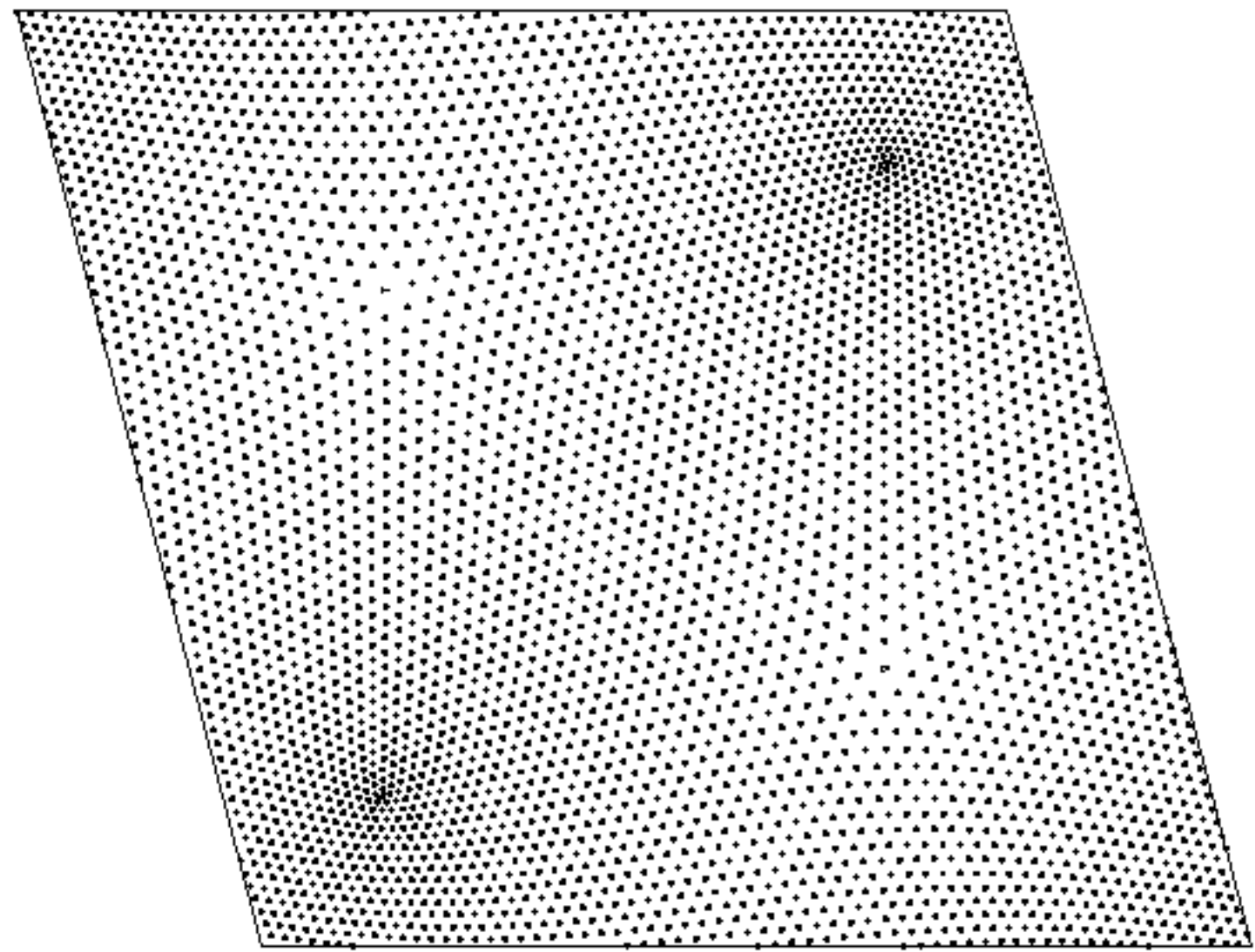
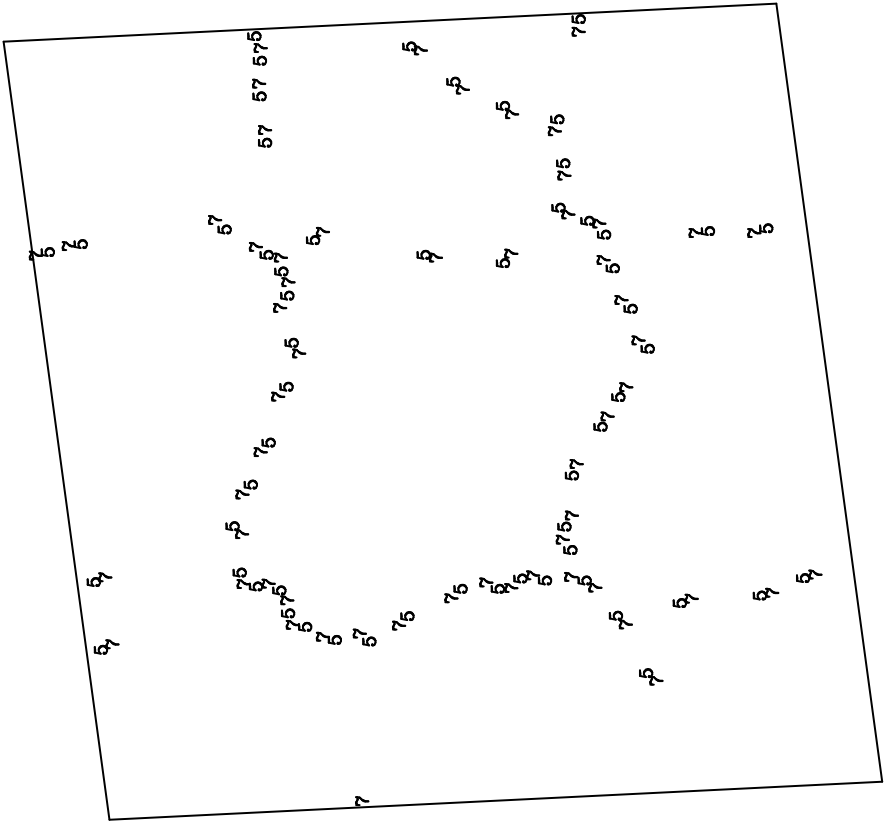
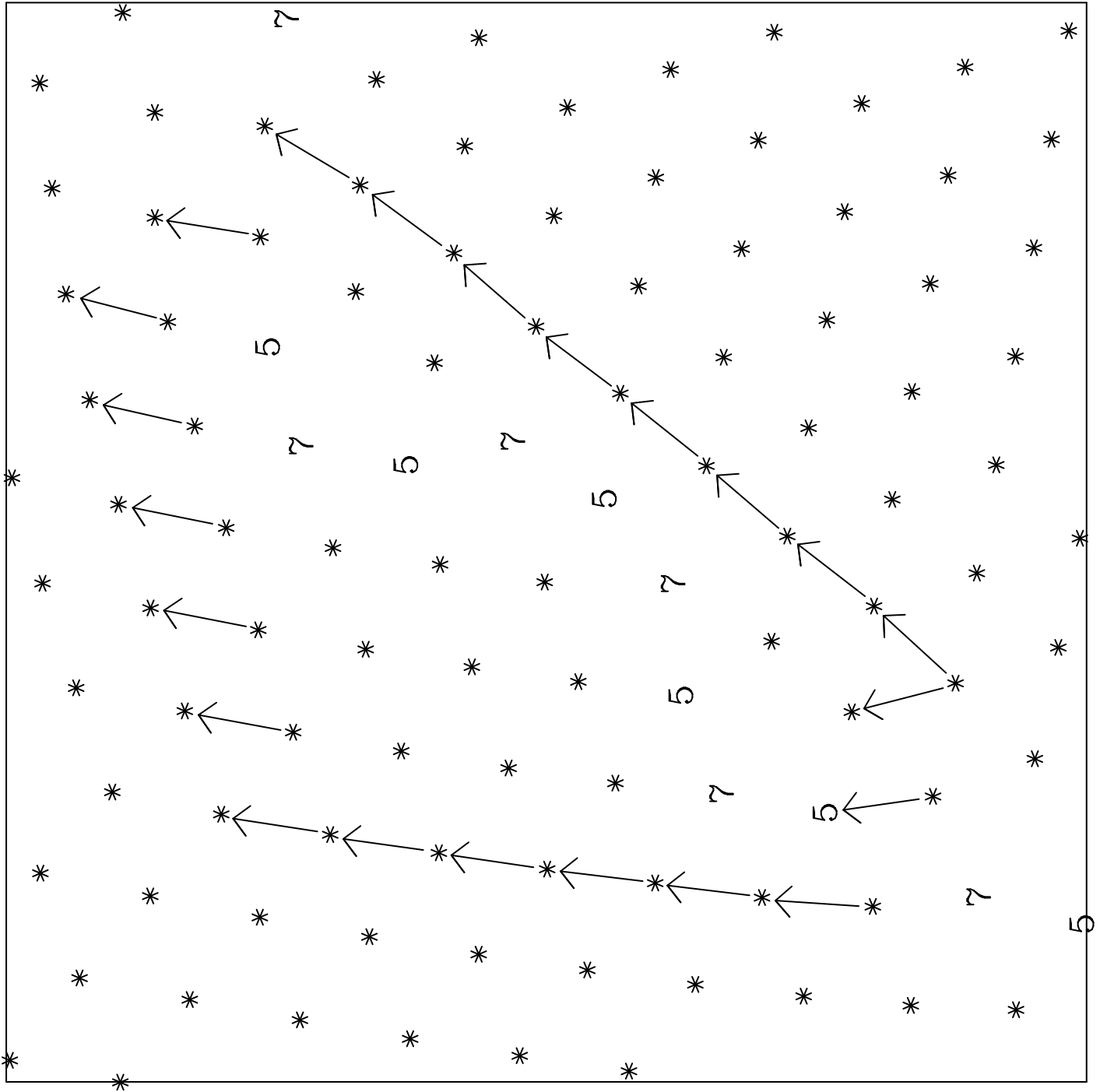
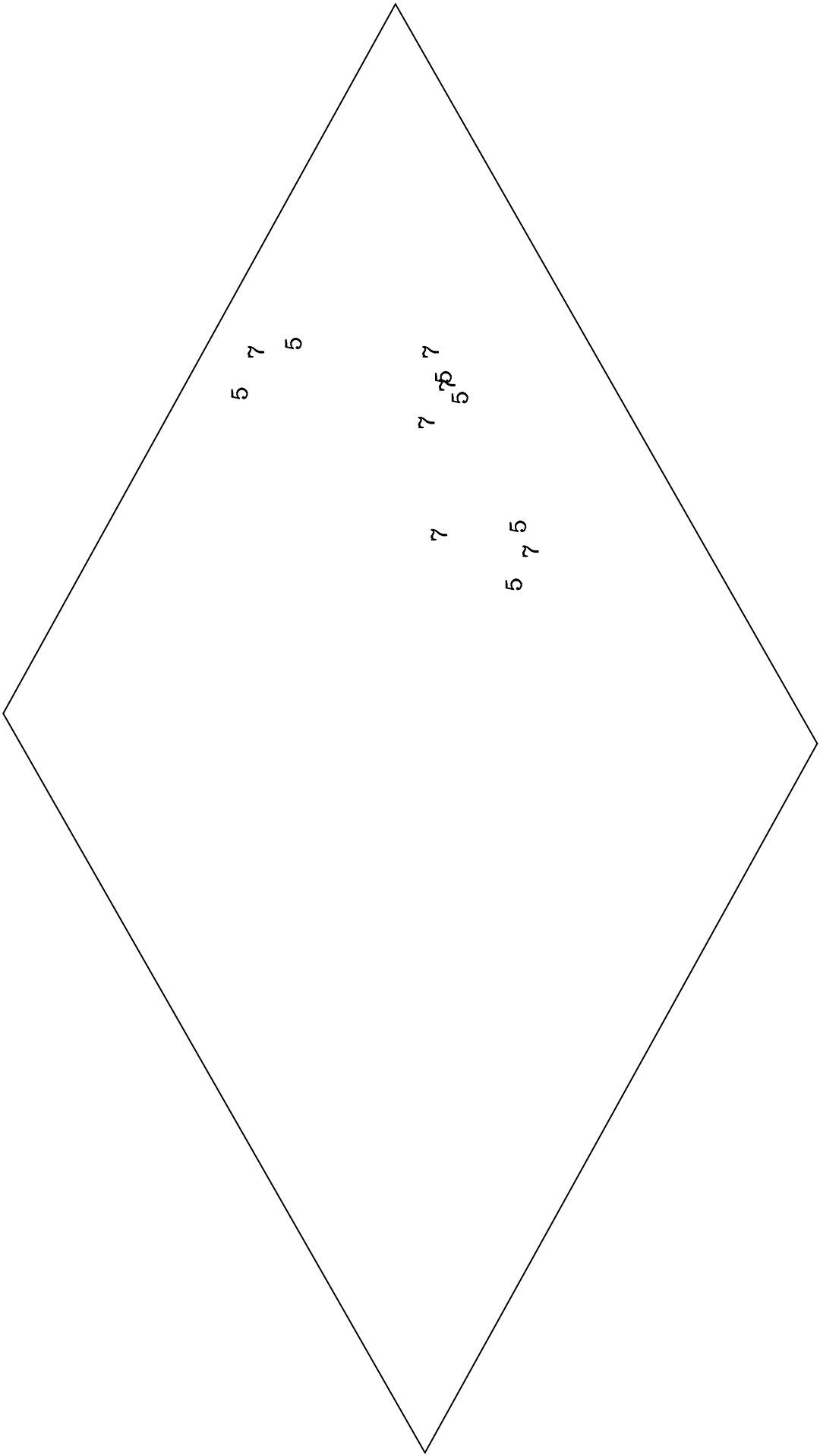
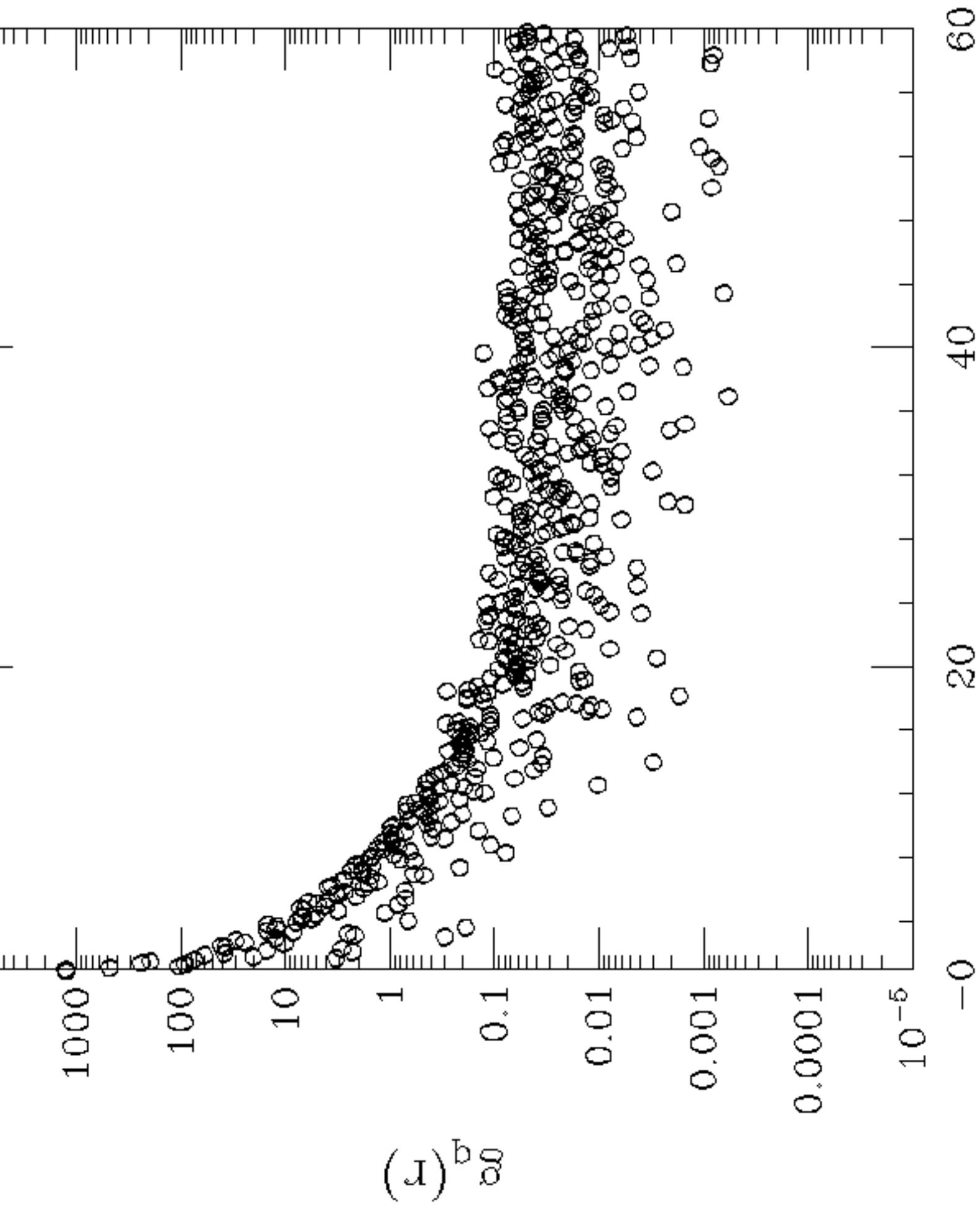
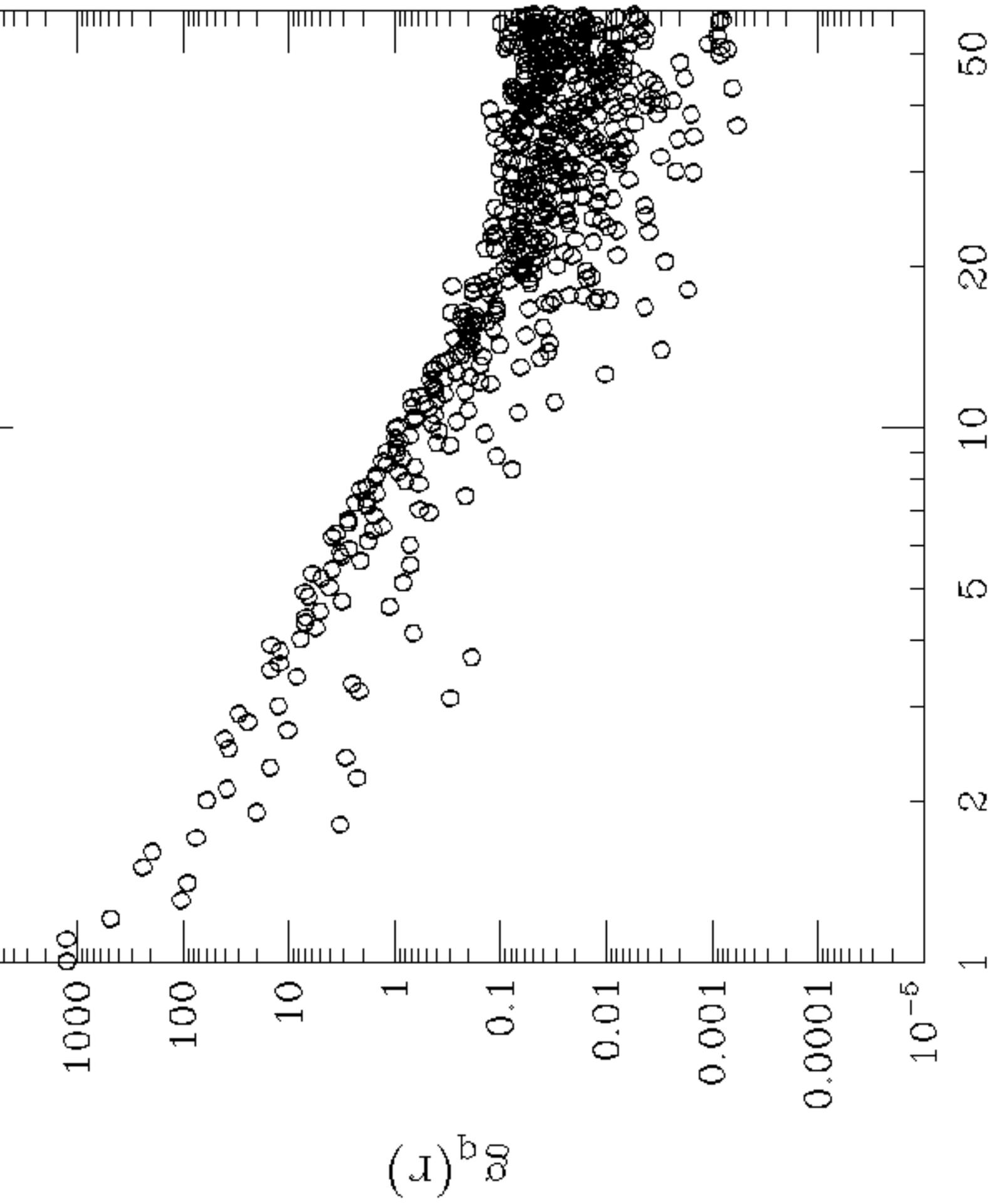
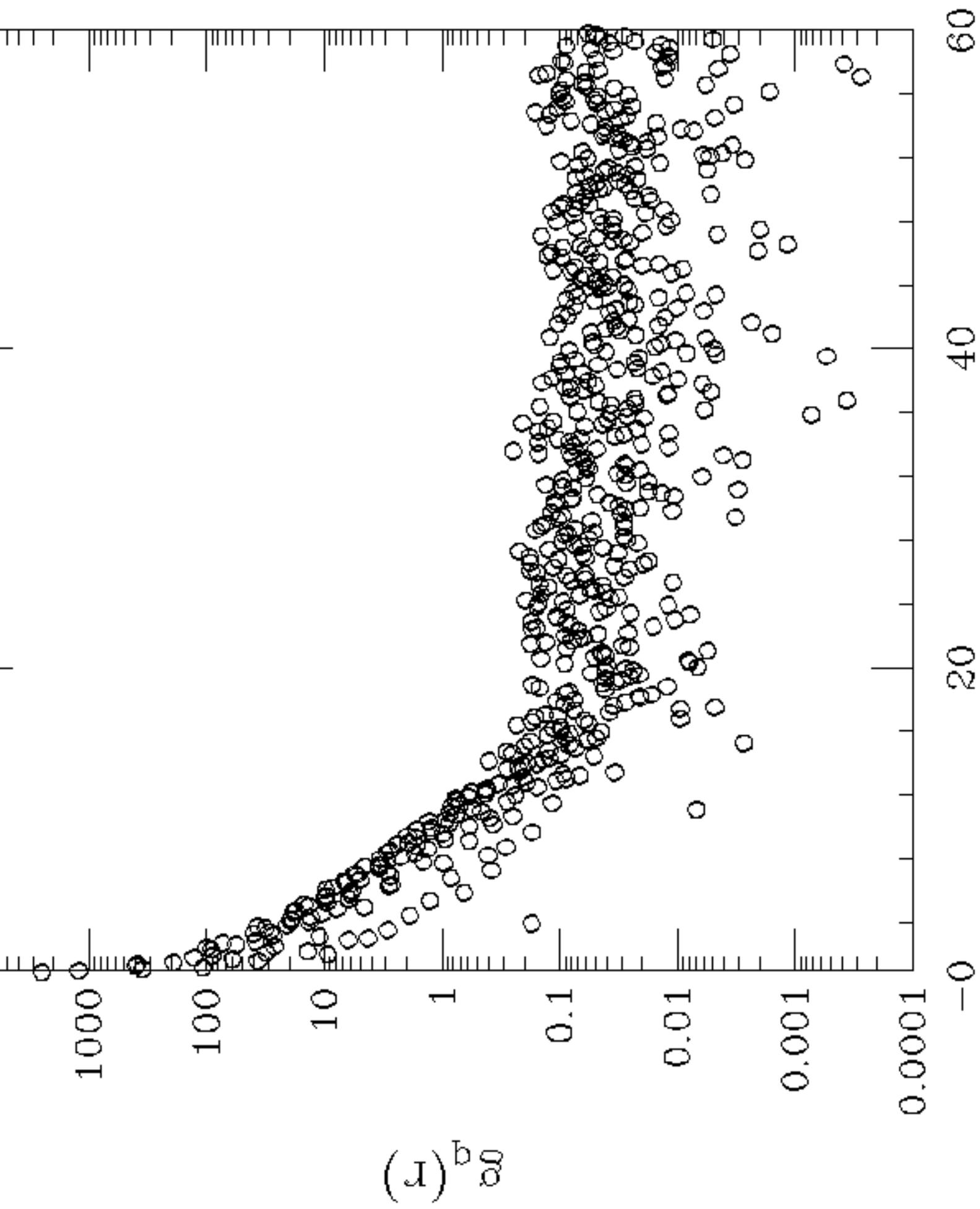
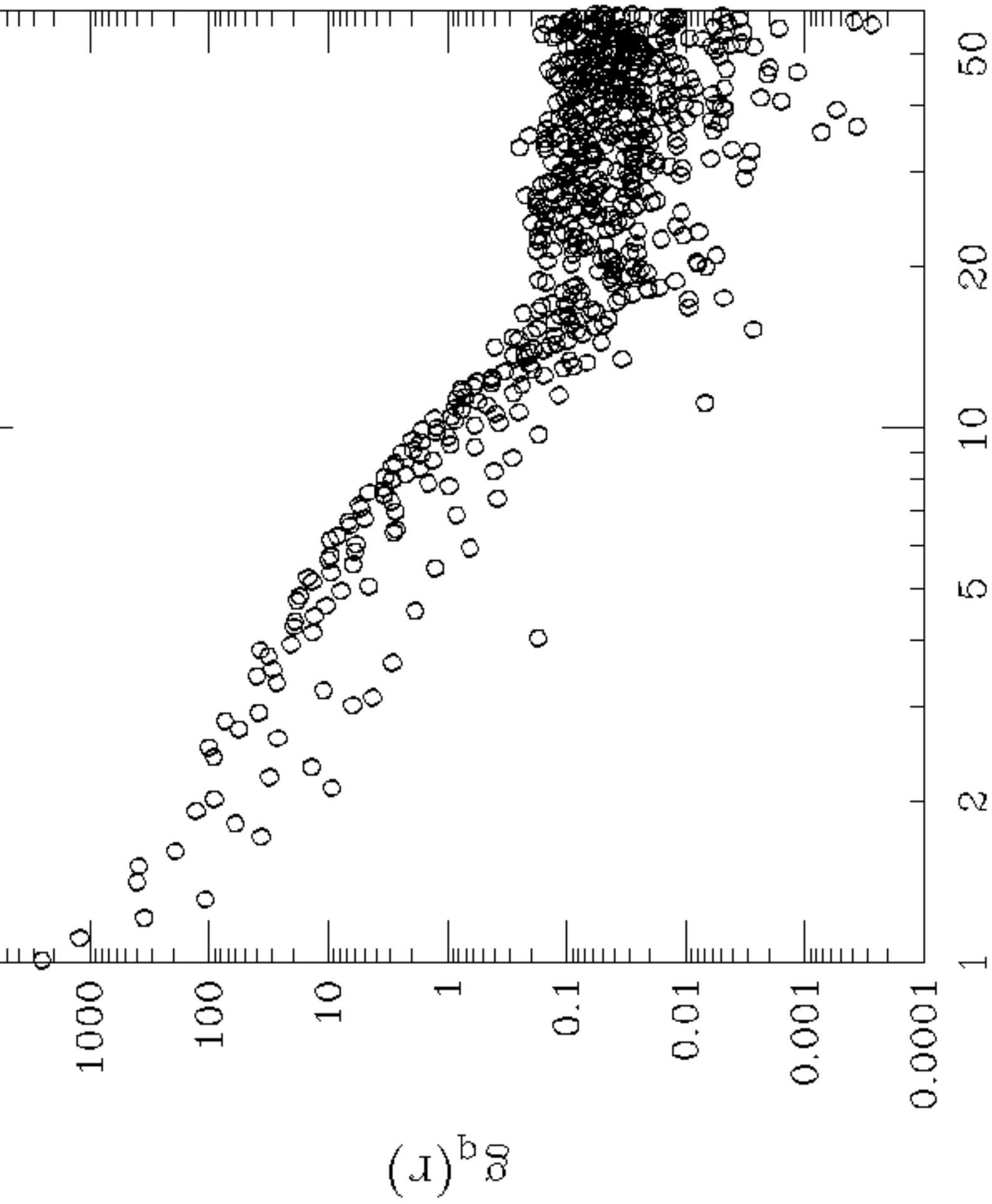
\end{document}